\documentclass[prl,aps,showpacs,superscriptaddress,twocolumn,floatfix]{revtex4-1}
\usepackage{dcolumn}
\usepackage{bm}
\usepackage{amsmath,amssymb,amsfonts,latexsym,fancyhdr,graphicx}

\newcommand{\ket}[1]{\displaystyle{|#1\rangle}}
\newcommand{\bra}[1]{\displaystyle{\langle#1|}}

\newcommand{\la}{\lambda}

\begin{document}
\title{Sudden transition between classical and quantum decoherence}
\author{L. Mazzola}\email{laura.mazzola@utu.fi}
\affiliation{Turku Centre for Quantum Physics, Department of Physics and Astronomy, University of
Turku, FI-20014 Turun yliopisto, Finland}
\author{J. Piilo}\email{jyrki.piilo@utu.fi}
\affiliation{Turku Centre for Quantum Physics, Department of Physics and Astronomy, University of
Turku, FI-20014 Turun yliopisto, Finland}
\author{S. Maniscalco}\email{sabrina.maniscalco@utu.fi}
\affiliation{Turku Centre for Quantum Physics, Department of Physics and Astronomy, University of
Turku, FI-20014 Turun yliopisto, Finland}

\begin{abstract}
We study the dynamics of quantum and classical correlations in the
presence of nondissipative decoherence. We discover a class of initial states for
which the quantum correlations, quantified by the quantum discord,
are not destroyed by decoherence for times $t < \bar{t}$. In this
initial time interval classical correlations decay. For  $t >
\bar{t}$, on the other hand, classical correlations do not change in
time and only quantum correlations are lost due to the interaction
with the environment. Therefore, at the transition time $\bar{t}$
the open system dynamics exhibits a sudden transition from classical to
quantum decoherence regime.
\end{abstract}

\maketitle

The interaction of a quantum system with its environment causes the
rapid destruction of crucial quantum properties, such as the
existence of quantum superpositions and of quantum correlations in
composite systems \cite{Breuer,Zurek}. Contrarily to the exponential
decay characterizing the transition from a quantum superposition to
the corresponding statistical mixture, entanglement may disappear
completely after a finite time, an effect known as entanglement
sudden death \cite{yu}. There exists, however, quantum correlations
more general and more fundamental than entanglement. Several
measures of these quantum correlations have been investigated in the
literature \cite{Henderson,Olliver,Opp,Grois,Luo,Modi}, and among
them the quantum discord \cite{Olliver,Henderson}, has recently received a great deal of attention
\cite{Zurek03,Horodec,Rodriguez,Datta,Piani,Maziero0,Shabani,Datta2,Piani2,Werlang,Maziero,Fanchini,Ferraro,Datta3}. \\
The total correlations (quantum and classical) in a bipartite
quantum system are measured by the quantum mutual information ${\cal
I}(\rho_{AB})$ defined as
\begin{equation}\label{mutual}
{\cal I}(\rho_{AB})=S(\rho_{A})+S(\rho_{B})-S(\rho_{AB})
\end{equation}
where $\rho_{A(B)}$ and $\rho_{AB}$ are the reduced density matrix
of subsystem $A(B)$ and the density matrix of the total system,
respectively, and $S(\rho)=-Tr\{\rho \log_{2}\rho\}$ is the von
Neumann entropy. The quantum discord is then defined as
\begin{equation}\label{discord}
{\cal D}(\rho_{AB})\equiv{\cal I}(\rho_{AB})-{\cal C}(\rho_{AB})
\end{equation}
where ${\cal{C}}(\rho_{AB})$ [see Eq. \eqref{classical}] are the classical correlations of the
state \cite{Henderson,Opp,Grois}. The quantum discord
measures quantum correlations of a more general type than
entanglement, there exists indeed separable mixed states having
nonzero discord \cite{Datta}. Interestingly, it has been proven both theoretically and
experimentally that such states provide computational speedup
compared to classical states in some quantum computation models
\cite{Datta,Lanyon}. \\
The dynamics of quantum and classical correlations in presence of
both Markovian \cite{Maziero0,Maziero} and non-Markovian
\cite{Fanchini} decoherence has been recently investigated.
It is believed that the quantum correlations measured by the quantum
discord, in the Markovian case, decay exponentially in time and
vanish only asymptotically \cite{Werlang,Ferraro}, contrarily to the
entanglement dynamics where sudden death may occur. \\
A remarkable result we demonstrate in this Letter is the existence of a class of initial states for which the quantum discord does
not decay for a finite time interval $0 < t < \bar{t}$ despite the
presence of a noisy environment. Our result is derived for qubits interacting with nondissipative independent reservoirs. It is not yet known whether such phenomenon can be observed for more general types of environment. However, this is the first evidence of the existence of quantum properties, in this case quantum correlations, that remain intact  under the action of an open quantum channel. \\
The major obstacle to the development of quantum technologies has been, until now, the destruction of all quantum properties caused by the inevitable interaction of quantum systems with their environment. The fact that, under certain conditions, quantum correlations useful for quantum algorithms are completely unaffected by the environment, for long time intervals, may constitute a new breakthrough to quantum technologies such as, e.g., quantum computers. \\
A crucial aspect of the dynamics is that, while the total quantum correlations measured by the discord remain constant, classical correlations are lost.
Interestingly, in this dynamical region, entanglement decays
exponentially in time but at the same time
quantum-correlations-other-than-entanglement, measured by dissonance \cite{Modi}, increase monotonically until $t = \bar{t}$. \\
We prove analytically that, for certain initial  Bell-diagonal states, when discord starts
to decay, i.e., for $t > \bar{t}$, the classical correlations become
constant in time. Therefore, there exists an instant of time
$\bar{t}$ at which the system stops losing classical correlation and
starts losing quantum correlations. The time $\bar{t}$ depends on a
single parameter characterizing the initial state. The class of
initial states for which the sudden transition from quantum to
classical decoherence occurs depends on the type of
Markovian noise considered. \\
Let us begin by specifying the quantity used for measuring the
classical correlations and, therefore, to calculate the quantum
discord by means of Eq.~(\ref{discord}). Such a quantity is in fact
a second extension of classical mutual information and it is based
on the generalization of the concept of conditional entropy. We know
that performing measurements on system $B$ affects our knowledge of
system $A$. How much system $A$ is modified by a measurement of $B$
depends on the type of measurement performed on $B$. Here the
measurement is considered of von Neumann type and it is described by
a complete set of orthonormal projectors $\{{\Pi_{k}}\}$ on
subsystem $B$  corresponding to the outcome $k$. The classical
correlations ${\cal C}(\rho_{AB})$ are then defined as
\cite{Henderson}
\begin{equation}\label{classical}
{\cal C}(\rho_{AB}) =  \max_{\{\Pi_{k}\}}[S(\rho_A)-S(\rho_{AB}|\{\Pi_{k}\})],
\end{equation}
where the maximum is taken over the set of projective measurements
$\{{\Pi_{k}}\}$ and
$S(\rho_{AB}|\{\Pi_{k}\})=\sum_{k}p_{k}S(\rho_{k})$ is the
conditional entropy of $A$, given the knowledge of the state of $B$,
with  $\rho_{k}=Tr_B (\Pi_{k} \rho_{AB} \Pi_{k})/p_k$ and
$p_k=Tr_{AB}(\rho_{AB} \Pi_{k})$. \\
We consider the case of two qubits under local nondissipative
channels, more specifically we focus on phase flip, bit flip and
bit-phase flip channels. For each qubit, the Markovian dissipator is given by ${\cal L}[\rho_{A(B)}]=\gamma [\sigma^{A(B)}_j \rho_{A(B)} \sigma_j^{A(B)} - \rho_{A(B)}]/2$, with  $\sigma_{j}^{A(B)}$  the Pauli operator in direction $j$ acting on $A (B)$, and $j=1,2,3$ for the bit, bit-phase, and phase flip cases, respectively. For simplicity, we take as initial states
of the composite system a class of states with maximally mixed
marginals
\begin{equation}
\rho_{AB}=\frac{1}{4} \left( \mathbf{ 1}_{AB} + \sum_{i=1}^3 c_i \sigma_{i}^{A}\sigma_{i}^{B}\right), \label{initial state}
\end{equation}
where $c_{i}$ is a real number such that
$0\leq|c_{i}|\leq1$ for every $i$ and $\mathbf{ 1}_{AB}$ the
identity operator of the total system. This class of states includes
the Werner states ($|c_1|=|c_2|=|c_3|=c$) and the Bell states
$|c_1|=|c_2|=|c_3|=1$. \\
We firstly focus on  the phase damping (or phase flip) channel. For the initial state of Eq.~(\ref{initial state}), the time evolution
of the total system is given by \cite{Maziero0}
\begin{eqnarray}
\rho_{AB}(t)&=&\la_{\Psi}^+(t)\ket{\Psi^{+}}\bra{\Psi^{+}}+\la_{\Phi}^+(t)\ket{\Phi^{+}}\bra{\Phi^{+}} \nonumber \\
&+&\la_{\Phi}^-(t)\ket{\Phi^{-}}\bra{\Phi^{-}}+\la_{\Psi}^-(t)\ket{\Psi^{-}}\bra{\Psi^{-}}, \label{evolution}
\end{eqnarray}
where
\begin{eqnarray}
\la_{\Psi}^{\pm}(t)&=&[1 \pm c_{1}(t) \mp c_{2}(t)+c_{3}(t)]/4, \label{la0} \\
\la_{\Phi}^{\pm}(t)&=&[1\pm c_{1}(t) \pm c_{2}(t)-c_{3}(t)]/4,\label{la3}
\end{eqnarray}
and $\ket{\Psi ^{\pm}}=(\ket{00}\pm\ket{11})/\sqrt{2}$,
$\ket{\Phi^{\pm}}=(\ket{01}\pm\ket{10})/\sqrt{2}$ are the four Bell
states. The time dependent coefficients in Eqs.
(\ref{la0})-(\ref{la3}) are $c_1(t)=c_1(0) \exp(- 2 \gamma t)$,
$c_2(t)=c_2(0) \exp(- 2 \gamma t)$, and $c_3(t)=c_3(0)\equiv c_3$,
with $\gamma$ the phase damping rate.\\
The mutual information ${\cal I}[\rho_{AB}(t)]$ and the classical
correlation ${\cal C}[\rho_{AB}(t)]$ in this case are given by \cite{Luo}
\begin{eqnarray}
&&\!\!\!\!\!\!\!\!\!{\cal I}[\rho_{AB}(t)]=2+\sum_{k,l}\lambda_{k}^l(t)\log_{2}\lambda_{k}^l(t), \label{mut} \\
&& \!\!\!\!\!\!\!\!\!{\cal C}[\rho_{AB}(t)]=\sum_{j=1}^{2}\frac{1+(-1)^{j}\chi(t)}{2}\log_{2}[1+(-1)^{j}\chi(t)], \label{cla}
\end{eqnarray}
where $\chi(t)=\max\{|c_{1}(t)|,|c_{2}(t)|,|c_{3}(t)|\}$, $k=\Psi,\Phi$, and $l=\pm$.
We note that the maximization procedure with respect to the projective measurements, present in the definition of the classical correlations of Eq.~(\ref{classical}), can be performed explicitly for the system here considered noticing that (i) the complete set of orthogonal projectors is given by $\Pi_j = \vert \theta _j \rangle \langle \theta_j \vert$, with $j=1,2$, $\vert \theta_{1} \rangle = \cos\theta \vert 0 \rangle + e^{i \phi} \sin \theta \vert 1 \rangle$, $\vert \theta_{2} \rangle = e^{-i \phi} \sin \theta \vert 0 \rangle - \cos\theta \vert 1 \rangle $; and (ii) the state of the system remain always of the form given by Eq.~(\ref{initial state}) during the time evolution.  \\
We now focus on the class of initial states for which $c_1(0)=\pm 1$
and $c_2(0)=\mp c_3(0)$, with $|c_3|< 1$. These states are mixtures
of Bell states of the form
\begin{equation}
\rho_{AB}=\frac{(1+c_{3})}{2}\ket{\Psi^{\pm}}\bra{\Psi^{\pm}}+\frac{(1-c_{3})}{2}\ket{\Phi^{\pm}}\bra{\Phi^{\pm}}. \label{eq:in}
\end{equation}
Inserting Eqs. (\ref{la0})-({\ref{la3}}) into Eq.~ (\ref{mut}) it is
straightforward to prove that, for this initial condition, the
mutual information takes the form
\begin{eqnarray}\label{main}
{\cal I}[\rho_{AB}(t)]&=&\sum_{j=1}^{2}\frac{1+(-1)^{j}
c_3}{2}\log_{2}[1+(-1)^{j}c_3]\\
&+&\sum_{j=1}^{2}\frac{1+(-1)^{j}
c_1(t)}{2}\log_{2}[1+(-1)^{j}c_1(t)] \nonumber.
\end{eqnarray}
Having in mind Eq.~(\ref{cla}) and remembering that $c_1(t)=\exp (- 2 \gamma t)$, one
sees immediately that, for $t < \bar{t}= - \ln (|c_3|)/(2\gamma)$,
the second term in Eq.~(\ref{main}) coincides with the classical
correlation  ${\cal C}[\rho_{AB}(t)]$, since $|c_{1}(t)|
> |c_{2}(t)|, |c_{3}(t)|=|c_{3}|$. The quantum discord is then given
by the first term of Eq.~(\ref{main}). Hence, for $t < \bar{t} $,
the quantum discord is constant in time. We note that, by changing
the initial condition, and in particular $|c_{3}|$, we can increase
the time interval $t < \bar{t}$ over which the discord is constant.
For increasing values of $\bar{t}$, however, the quantum discord
decreases towards its zero value obtained for $|c_{3}|=0$. \\
In Fig.~1 we plot the time evolution of the quantum discord,
the classical correlations and the mutual information for
$c_1(0)=1$, $c_2(0)=-c_3$ and $c_3=0.6$.
\begin{figure}
\includegraphics[scale=0.6]{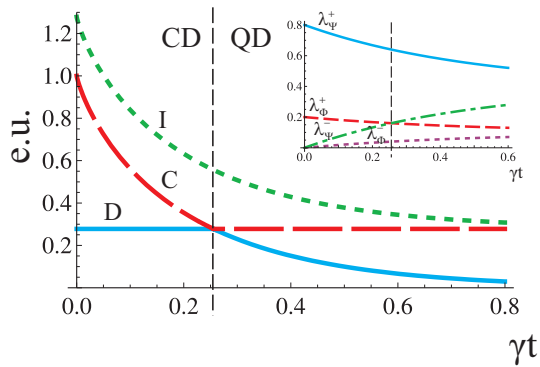} \caption{(Colors online)
Dynamics of mutual information (green dotted line), classical
correlations (red dashed line) and quantum discord (blue solid line)
as a function of $\gamma t$ for $c_1(0)=1$, $c_2(0)=-c_3$ and
$c_3=0.6$. In the inset we plot the eigenvalues $\lambda_{\Psi}^{+}$ (blue
solid line), $\lambda_{\Psi}^{-}$ (green dash-dotted line),
$\lambda_{\Phi}^{+}$ (red dashed line) and $\lambda_{\Phi}^{-}$ (violet
dotted line) as a function of $\gamma t$ for the same parameters.}
\end{figure}
The plot clearly shows the sharp transition from the classical to
the quantum decoherence regime occurring at $t=\bar{t}$. \\
In order to understand the physical origin of the sudden transition from classical to quantum decoherence, we consider the distances between our state and (i) its closest classical state and  (ii) its closest separable state. We adopt the definitions proposed in Ref.~\cite{Modi}, and we measure all distances by means of the relative entropy. In this way, the former distance coincide with a second definition of discord, while the latter distance is the relative entropy of entanglement. We begin by demonstrating that, for the system here considered, the discord defined by Eq.~(\ref{discord}) coincides with the one introduced in Ref.~\cite{Modi}. To this aim we notice that, the classical state closest to the state of our system at time $t$, given by Eq.~(\ref{evolution}), is \cite{Modi}
\begin{equation}
\rho_{\rm cl}(t)=\frac{q(t)}{2} \sum_{i=1,2} \ket{\Psi_i}\bra{\Psi_i}+\frac{1-q(t)}{2} \sum_{i=3,4}\ket{\Psi_i}\bra{\Psi_i}, \label{closclas}
\end{equation}
with $q(t)=\lambda_1(t)+\lambda_2(t)$, where $\lambda_1(t)$ and $\lambda_2(t)$ are the two highest eigenvalues given by Eqs.~(\ref{la0})-(\ref{la3}), and $\ket{\Psi_i}$ the corresponding Bell states. In the inset of Fig.~1 we plot the eigenvalues $\lambda_{\Psi}^{\pm}$ and  $\lambda_{\Phi}^{\pm}$, giving the weights or populations of the four Bell states components. The inset shows that at $t={\bar t}$ the population of $\ket{\Phi_+}$ becomes equal to the population of $\ket{\Psi_-}$ and, subsequently, it continues to decrease while the other one grows. As a consequence of this switch in the second highest population component, for $t < {\bar t}$,  the closest classical state is
\begin{eqnarray}
\rho_{\rm cl}(t<\bar{t})&=&\frac{1+e^{-2 \gamma t}}{4}\left(\ket{\Psi^{+}}\bra{\Psi^{+}}+ \ket{\Phi^{+}}\bra{\Phi^{+}} \right) \nonumber \\
&+&\frac{1-e^{-2 \gamma t}}{4}\left( \ket{\Phi^{-}}\bra{\Phi^{-}}+\ket{\Psi^{-}}\bra{\Psi^{-}} \right), \label{closclasC}
\end{eqnarray}
while, for $t > {\bar t}$,
\begin{eqnarray}
\rho_{\rm cl}(t>\bar{t})&=&\frac{1+c_3}{4}\left(\ket{\Psi^{+}}\bra{\Psi^{+}}+ \ket{\Psi^{-}}\bra{\Psi^{-}} \right) \nonumber \\
&+&\frac{1-c_3}{4}\left( \ket{\Phi^{-}}\bra{\Phi^{-}}+\ket{\Phi^{+}}\bra{\Phi^{+}} \right). \label{closclasQ}
\end{eqnarray}
Let us now look at the dynamics of the relative entropy $D(\rho_{AB} \| \rho_{\rm cl})=-Tr \{\rho_{AB} \log_2 \rho_{\rm cl}\} + Tr \{\rho_{AB} \log \rho_{AB}\}$ \cite{Modi}.
Inserting Eq.~(\ref{evolution}) and Eqs.~(\ref{closclasC})-(\ref{closclasQ}) into the expression for $D(\rho_{AB} \| \rho_{\rm cl})$, it is straightforward to prove that $D(\rho_{AB} \| \rho_{\rm cl})={\cal D}(\rho_{AB})$. This result holds for all the states of the form of Eq.~(\ref{initial state}). Hence, in the first dynamical regime, when the discord is constant and only classical correlations are lost, the distance to the closest classical state remains constant. At $t={\bar t}$, the closest classical state changes suddenly from the one given by Eq.~ (\ref{closclasC}) to the one given by Eq.~(\ref{closclasQ}). Subsequently, for $t>{\bar t}$, $\rho_{\rm cl}$ remains constant in time and the state of the system approaches asymptotically such state, as indicated by the monotonic decay of the quantum discord.
This behavior suggests a sufficient condition for the occurrence of  the sudden transition between classical and quantum decoherence. This transition  is present in the dynamics for those classes of initial states and dynamical maps for which (i) the state is at all times of the form of Eq.~(\ref{initial state}) and (ii) its distance to the closest classical state is constant. \\
To further understand the dynamics of the total quantum correlations, we study the relative entropy of entanglement $E$ and the dissonance $Q$ defined as the distance to the closest separable state  $\rho_S$ and the distance between  $\rho_S$ and its closest classical state  $\rho_{SC}$, respectively \cite{Modi}. Both quantities can be calculated exactly in our model. Entanglement takes the simple form $E=1+\lambda_1 \log_2 \lambda_1 + (1-\lambda_1) \log_2 (1-\lambda_1)$, with $\lambda_1$ the highest of the eigenvalues given by Eqs. (\ref{la0})-(\ref{la3}). This equation shows that entanglement always decays monotonically and it vanishes completely for $t \ge t_S=-\ln[(1-|c_3|)/(1+|c_3|)]/(2\gamma)$. This result is independent of the entanglement measure since all entanglement measures coincide and are equal to zero for separable states. If $t_S < \bar{t}$, then
entanglement disappears completely when the quantum discord has not
yet started to decay so the state of the total system is a separable
state with nonzero discord. These are the states exploited in the
one-qubit model of quantum computation of Ref.~\cite{Datta}. One can
easily check that $t_S < \bar{t}$ whenever $0<|c_3|<\sqrt{2}-1$. Figure 2 shows one of such examples. Moreover, there exist classes of initial separable states for which discord remains constant for $t < {\bar t}$ while entanglement is always zero. It is simple to see by direct substitution that, e.g., the state of the form of Eq.~ (\ref{initial state}) with $c_1=\pm (1-|c_3|)/(1+|c_3|)$, $c_2=-c_3  (1-|c_3|)/(1+|c_3|)$, and $0<|c_3|<\sqrt{2}-1$  displays this behavior.\\
Let us, finally, look at the dissonance. We obtain $Q=1+\sum_{i=1}^4 p_i \log_2 p_i-(p_1+p_2) \log_2 (p_1+p_2)+(1-p_1-p_2)\log_2 (1-p_i-p_2)$, with $p_1=1/2$, $p_i=\lambda_i/2(1-\lambda_1)$, and $\lambda_1\ge \lambda_2 \ge \lambda_3\ge \lambda_4$ the eigenvalues of Eqs. (\ref{la0})-(\ref{la3}) in non increasing order.
Figure 3 shows the time evolution of discord, entanglement and dissonance, all measured in entropic units. Remarkably, while entanglement decays, dissonance increases monotonically in time until $t={\bar t}$. This means that while the state of the system approaches its closest separable state, this state in turns goes farther and farther from its closest classical state. The increase in the dissonance indicates an increase in other-than-entanglement quantum correlations which contribute to mantain the total quantum correlations (discord) constant. It is worth noticing, however, that, as noted in Ref. \cite{Modi}, dissonance and entanglement do not add to give the discord because of the subadditivity of correlations.
It is simple to see that, for the bit flip and phase-bit flip channels, the class of states for which the sudden transition from classical to quantum decoherence occurs, have the same form of Eq.~(\ref{eq:in}), with $c_1$ and $c_2$ replacing $c_3$, respectively.
\begin{figure}
\includegraphics[scale=0.5]{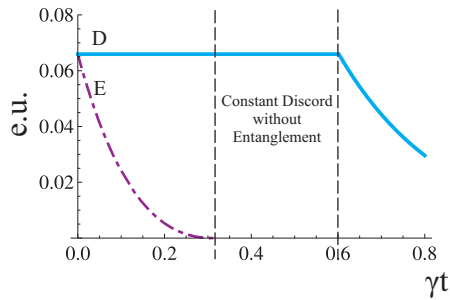} \caption{(Colors online)
Dynamics of entanglement (violet dashed-dotted line) and quantum discord (blue solid line)
as a function of $\gamma t$ for $c_1(0)=1$, $c_2(0)=-c_3$ and
$c_3=0.3$. }
\end{figure}
\begin{figure}
\includegraphics[scale=0.5]{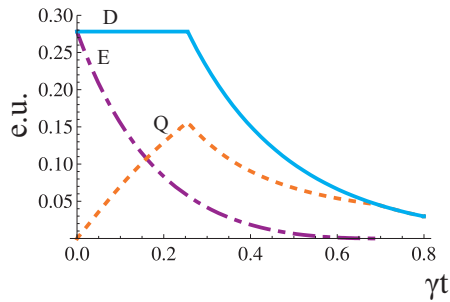} \caption{Dynamics of entanglement (violet dashed-dotted line), quantum discord (blue solid line) and dissonance (orange dashed line)
as a function of $\gamma t$ for $c_1(0)=1$, $c_2(0)=-c_3$ and
$c_3=0.6$.}
\end{figure}
\\The existence of a sharp transition between classical and quantum
loss of correlations in a composite system is a remarkable feature
of the dynamics of composite open quantum system that was up to now
unknown. The existence of a finite time interval during which
quantum correlations initially present in the state do not decay in
presence of decoherence opens a series of interesting questions. Is
it possible to exploit the class of initial states displaying such a
property to perform quantum computation or communication tasks
without any disturbance from the noisy environment for long enough
intervals of time? Which is the most general class of states and of open quantum systems
exhibiting a sudden transition from classical to quantum decoherence?
Finally, and perhaps most importantly, which are the physical
mechanisms that forbid the loss of quantum correlations at the
initial times and that allow only quantum correlations to be lost
after the transition time $\bar{t}$? We believe that the transition from classical to quantum decoherence presented in this Letter and very recently confirmed experimentally \cite{Jin-Shi} will shed new light on one of the most fundamental
and fascinating aspects of quantum theory.

This work has been supported by the Academy of Finland (Project No.~133682), the Magnus
Ehrnrooth Foundation, the Emil Aaltonen Foundation, and the Finnish Cultural Foundation. S.M. also thanks
the Turku Collegium of Science and Medicine for financial support. L.M. acknowledges K. Modi for useful discussions.

\end{document}